\documentclass[%
 reprint,
superscriptaddress,
 amsmath,amssymb,
 aps,
pra
]{revtex4-2}

\usepackage{xcolor}
\usepackage{graphicx}
\usepackage{dcolumn}
\usepackage{bm}

\begin{document}
\title{Emission engineering in monolithically integrated silicon nitride microring resonators}

\author{Kishor Kumar Mandal}
\thanks{These two authors contributed equally}

\author{Anuj Kumar Singh}
\thanks{These two authors contributed equally}

\affiliation{%
Laboratory of Optics of Quantum Materials, Department of Physics, Indian Institute of Technology Bombay, Mumbai- 400076, India 
}%

\author{Brijesh Kumar}
\affiliation{%
Laboratory of Optics of Quantum Materials, Department of Physics, Indian Institute of Technology Bombay, Mumbai- 400076, India 
}%
\author{Amit P. Shah}
\affiliation{Department of Condensed Matter Physics and Material Science, Tata Institute of Fundamental Research, Homi Bhabha Road, Mumbai, 400005 India}

\author{Rishabh Vij}
\affiliation{Department of Condensed Matter Physics and Material Science, Tata Institute of Fundamental Research, Homi Bhabha Road, Mumbai, 400005 India}

\author{Amrita Majumder}
\affiliation{%
Laboratory of Optics of Quantum Materials, Department of Physics, Indian Institute of Technology Bombay, Mumbai- 400076, India 
}
\author{Janhavi Jayawant Khunte}
\affiliation{%
Laboratory of Optics of Quantum Materials, Department of Physics, Indian Institute of Technology Bombay, Mumbai- 400076, India 
}

\author{Venu Gopal Achanta}
\email{achanta@tifr.res.in}
\affiliation{Department of Condensed Matter Physics and Material Science, Tata Institute of Fundamental Research, Homi Bhabha Road, Mumbai, 400005 India}
\author{Anshuman Kumar}%
 \email{anshuman.kumar@iitb.ac.in}
\affiliation{%
Laboratory of Optics of Quantum Materials, Department of Physics, Indian Institute of Technology Bombay, Mumbai- 400076, India 
}%
\affiliation{Centre of Excellence in Quantum Information, Computation, Science and Technology, Indian Institute of Technology Bombay, Powai, Mumbai 400076, India}

\date{\today}
\begin{abstract}
Monolithic integration of solid-state emitters with photonic elements of the same material is a promising approach to overcome the constraints of fabrication complexity and coupling losses in traditional hybrid integration approaches. 
 A wide band-gap, low-loss silicon nitride (SiN) platform is a mature technology, having CMOS compatibility, widely used in hybrid integrated photonics and optoelectronics. However, it has been shown that certain growth conditions enable the SiN material to host color centers, whose origin is currently under investigation. In this work, we have engineered a novel technique for the efficient coupling of these intrinsic emitters into the whispering gallery modes (WGMs) of the SiN microring cavity -- which has not been explored previously. We have engineered a subwavelength-sized notch into the rim of the SiN microring structure, to optimize the collection efficiency of the cavity-coupled enhanced photoluminescence (PL) spectra at room temperature. The platform presented in this work will enable the development of monolithic integration of color centers with nanophotonic elements for application to quantum photonic technologies.
\end{abstract}

\maketitle
\section{Introduction}
Rapid advancement in integrated photonic circuitry platforms on the industry scale has prompted tremendous progress to meet the demands for photonic technology. The hybrid photonic integration approach increases the complexity level of fabrication. Additionally, with hybrid integration, quantum photonic integrated circuits (QPICs) face significant challenges such as scalability, stability, optical losses, and low coupling efficiency of emitters due to misalignment and a heterogeneous material interface in the system, hence leading to degradation of quantum optical properties \cite{sasani2021chip, bogdanov2017material}. To address the challenges linked to heterogeneous material architecture, the development of alternative monolithic platforms is sought. This includes exploring intrinsic emitters within the same material platform, facilitating efficient photon routing with low losses in the QPICs platform at a large scale with a simple fabrication process flow. Silicon carbide (SiC) \cite{castelletto2020silicon, wolfowicz2020vanadium}, gallium nitride (GaN) \cite{nguyen2021site}, aluminum nitride (AlN) \cite{bishop2020room}, silicon nitride (SiN) \cite{senichev2021room, senichev2022silicon, smith2020single}, and silicon\cite{khoury2022bright} serve as wide-bandgap semiconductor optical materials, which host intrinsic quantum emitters (QEs), showcasing their potential as high-purity, bright, and stable single-photon emitters at room temperature.\\
Presently, silicon nitride has emerged as a promising contender, proving to be well-suited for on-chip integrated photonics through the monolithic construction of various types of photonic elements on the same platform. In addition to a remarkable single photon emitter source,  silicon nitride is a wide-bandgap semiconductor material (band gap tuning depends on deposition conditions \cite{kruckel2017optical, bucio2019silicon}) that also exhibits exceptional properties-- transparency window from visible to near-infrared wavelength regime \cite{subramanian2013low, kruckel2017optical}, thermo-optic response \cite{nejadriahi2020thermo,  bucio2019silicon}, non-linearity \cite{Ikeda2008}, negligible two-photon absorption at C-band \cite{kruckel2017optical}, low propagation and insertion loss, CMOS-compatibility, and fabrication flexibility \cite{moss2013new, romero2013silicon}. This makes SiN a technology-driven material that has a refractive index (n~$\sim$2) and offers relatively good index contrast ($\Delta$n) with underlying SiO$_{2}$ clad layer for efficient mode confinement with ultra-low propagation loss in compact and complex integration of components in a photonic chip. Leading photonics foundries-- LigenTec, Imec, LioniX, \cite{blumenthal2018silicon, siew2021review, munoz2019foundry} and quantum technology companies, for example, Xanadu \cite{arrazola2021quantum} and QuiX \cite{taballione202320} have adapted the SiN optical material platform for next-generation photonic or QPIC development.\\
\begin{figure*}
   \centering
   \includegraphics[width= 0.97\textwidth]{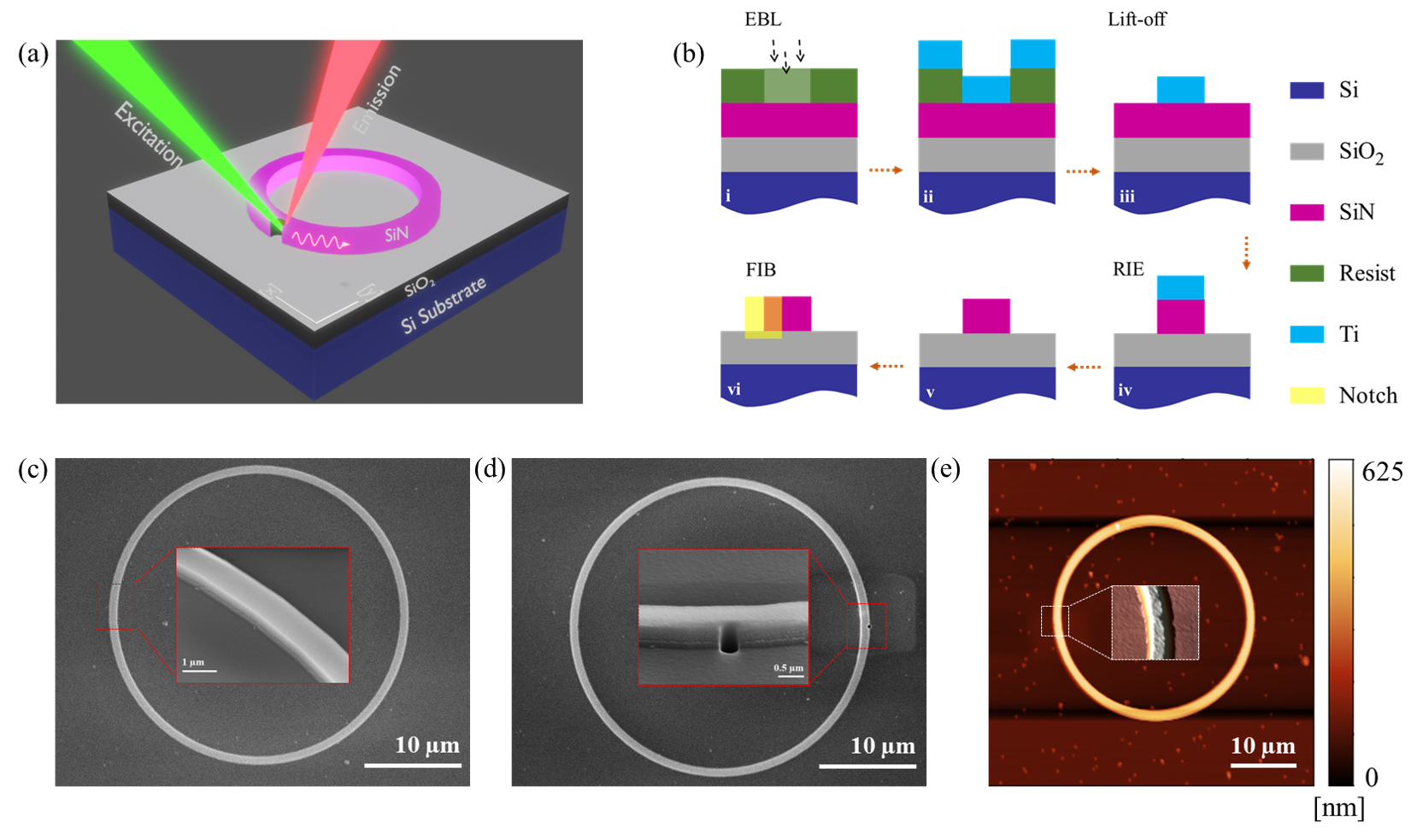}
    {\caption{\textbf{Nanophotonic structure fabrication using Si-photonics foundry process to study intrinsic quantum emitter in the SiN microring resonator}. (a) Schematic layout of coupled excitation and detection of cavity mode. A subwavelength size notch ($\sim\lambda/3$) design was introduced at the edge of the microring cavity structure to effectively coupling-in/out of light signals. (b) The fabrication process flows to a pattern notch-integrated SiN-microring cavity atop of SiO$_{2}$ platform. SEM micrograph of as-fabricated microcavity structure with (c) and without (d) notch at the rim of a cavity. The inset shows the enlarged view (in a tilted SEM mode) of the notch introduced. (e) Shows an AFM image of the microring without the notch, the inset depicts the site for the roughness estimation.}}
    \label{fig:figure5}
\end{figure*}
Engineered optical microcavities have attracted enormous attention to enhance light-matter interaction and have opened opportunities for in-depth fundamental research and its potential applications in the wide area of photonics and optoelectronics. Various configurations based on distinct photonic mode confinement physics of optical microcavities have been explored \cite{vahala2003optical, javerzac2018excitonic, eswaramoorthy2022engineering, singh2023low}. One popular class of cavities among them is the whispering gallery mode (WGM) type microcavity, which has shown high performance and on-chip integration compatibility \cite{du2020nanolasers, hepp2019semiconductor, yang2015advances, wang20192d, foreman2015whispering, chiasera2010spherical, bogaerts2012silicon, eswaramoorthy2022engineering, javerzac2018excitonic}. WGM microcavities offer a high Q factor and electromagnetic field localization in small mode volume. In the field of silicon photonics, the microring-like stable planar structure (crucial for semiconductor processing) is a commonly employed WGM-type microcavity.
Notably, efficient coupling-out of these trapped WGM cavity modes for out-of-plane or off-chip collection has encountered key challenges \cite{reed2015wavelength, lee2023shape}, unlike in-plane coupling scenarios (on-chip collection), where the cavity mode is routed by introducing a planar waveguide in close proximity through an evanescent field coupling at the coupling gap \cite{bogaerts2012silicon, feng2012silicon, van2016optical, hodaei2014parity}. Although several nanophotonic microcavity architectures have been investigated to obtain WGMs spectrum in off-chip collection schemes for example --- III-V microcavity structures\cite{wong2021epitaxially, guo2019recent}, and 2D materials\cite{du2020nanolasers}, they either employ non-planar geometries, require hybrid and complex integration or are not compatible with CMOS fabrication technology. Although extensive examination has been conducted on the photoluminescence (PL) in the context of SiN films and waveguides\cite{senichev2021room, senichev2022silicon, smith2020single, lan2023visualization, parkhomenko2017origin, kistner2011photoluminescence, assad2015two}, the exploration of the resonant modes of SiN microcavity, specifically the detection of whispering gallery modes (WGMs), through the excitation and coupling of a quantum light source embedded in SiN material, has not been investigated to the best of our knowledge. This underscores the primary goal of the present study. \\
In this work, we present an innovative approach to engineer WGM-type microcavity structures and showcase successful demonstrations of intrinsic quantum emitter-based whispering gallery modes within the CMOS-compatible planar SiN microring resonator nanophotonic architecture platform. We have designed a subwavelength notch (of size $\sim\lambda$/3) \cite{zhu2021design, arbabi2015grating, zhu2022research, chen2021photoluminescence, reed2016photothermal, reed2015wavelength} to create an artificial defect at the edge of the 30-$\mu$m-dia SiN microring structure. The notch acts both as an antenna for efficiently coupling laser light into the cavity to efficiently excite emitters in the SiN matrix and for coupling-out the emission into the cavity mode generated by the re-circulation of trapped light. This study has been carried out in a $\mu$-PL setup for capturing cavity-coupled PL spectrum from 570 to 870 nm wavelength span with co-located excitation and collection spots. 
We present a detailed analysis of the SiN film growth and its characterization to gain deeper insight into the nature of the embedded emitters in SiN film as well as the microcavity. The proposed SiN microring cavity integrated with a notch platform was fabricated using silicon photonics chip fabrication technology in a clean room environment. Mainly electron beam lithography, reactive ion etching, and focused Ga$^{+3}$-ion beam milling were employed to define cavity structure. The structures were designed by utilizing commercially available Ansys/Lumerical simulation software. We have used a finite difference eigenmode (FDE) solver for modal analysis and a finite-difference time-domain (FDTD) solution for WGMs cavity electric field contour mapping in both with and without a notch in microring cavity.

\section{Methods}
\subsection{SiN material growth}
 Intrinsic solid-state emitters in dielectric silicon nitride (SiN) material were prepared during a plasma-enhanced chemical vapour deposition (PECVD), which typically allows for thin film deposition at a lower temperature in comparison to other deposition systems. 
In the deposition process, the mixture of precursor gases flux, chamber pressure, temperature, RF power, and deposition rate are the key parameters for depositing high-quality film at a controllable rate within a plasma chamber in a capacitively-coupled PECVD tool operating at a radio frequency of 13.56 MHz. Prior to growth, a commercially available 500-$\mu$m thick Si-wafer (100) was cleaned by a standard RCA cleaning process to remove all types of organic residue and contamination particles, then a 2-$\mu$m thick SiO$_{2}$ layer was grown on Si-substrate as a bottom cladding layer to maintain optical isolation and also making suitable for SiN microcavity fabrication. With appropriate adjustment of precursor gases flux ratio NH${_3}$/SiH${_4}$ in addition to Ar gas flow in a small fraction, a 300-nm-thin SiN layer was deposited atop SiO$_{2}$ at a slow deposition rate. During the deposition, RF power, the chamber pressure, and the base plate template were maintained at 400 W, 5 Pa, and $200~ ^{\circ}$C, respectively.\\
\subsection{Fabrication of notch-integrated SiN-microring}
The realization of nanoscale features in integrated photonic chips typically requires an electron beam lithography (EBL) system. Here, we have employed a top-down approach to pattern the proposed SiN notch-integrated microring resonator to examine the cavity mode of the trapped intrinsic quantum emitter, as depicted in Fig. 1(a). The step-by-step fabrication flow in the cross-section view is sketched in Fig. 1(b) and the process was carried out by utilizing the CMOS photonic foundry technology. After the SiN deposition in PECVD, the following crucial steps have been executed in the process flow: writing by e-beam lithography, metallization by e-beam evaporator (EBE), dry etching by reactive ion etching (RIE), scanning electron microscope (SEM), and notch design by focused ion beam (FIB) milling. \\
Prior to the fabrication process, a diced sample was pre-cleaned with DI water, acetone, IPA, and DI water in a sonication bath followed by drying with a nitrogen gun and kept in a hotplate at 200$^{\circ}$C for dehydration. Subsequently, it was surface treated with a UV ozone cleaner (Ossila). Then a bi-layer polymethyl methacrylate (PMMA) e-beam resist of different sensitivities was spin-coated at 4000 r.p.m. and prebaked for 2 mins at 180$^{\circ}$C. The microcavity pattern was defined with EBL (RAITH 150$^{TWO}$) on polymer PMMA resist through a conventional area mode of writing scheme. The set EBL exposure parameters were: working distance (8mm), aperture size (10~$\mu$m), EHT (20~kV), beam current (0.052 nA), curve area dose (400 C/cm$^{2}$) and writing at 2000x magnification in a 50~$\mu$m write field area. Then, exposed samples were developed in a developer solution, MIBK/IPA 1:3 for 20 sec. A 75-nm-thin layer of titanium (Ti) metal was deposited in the exposed area using the EBE tool and afterward, the sample was kept in a remover PG bath for a longer time to lift-off resist. The defined Ti structure on the SiN layer acts as a hard mask to preserve high selectivity in the dry etching of the 300-nm-SiN layer by the RIE plasma chamber \cite{senichev2022silicon}. The structure was over-etched to a total height of $\sim$400 nm (including SiO$_{2}$ of $\sim$100 nm thickness), to completely remove the surface effect of SiN material and to only have the microring contribution. Fluorine-based gas chemistry (CHF$_{3}$/O$_{2}$) was used to etch out a dielectric SiN layer at RF power 150 W and chamber pressure 1 Pa. Finally, the Ti etch mask was removed with a chemical etchant solution (NH$_{4}$OH: H$_{2}$O$_{2}$) and rinsed in DI water. The SEM micrograph of the as-fabricated microring cavity is shown in Fig. 1(c) and the insets portray the magnified view of important features. \\\\
\begin{figure*}
   \centering
   \includegraphics[width= 0.99\textwidth]{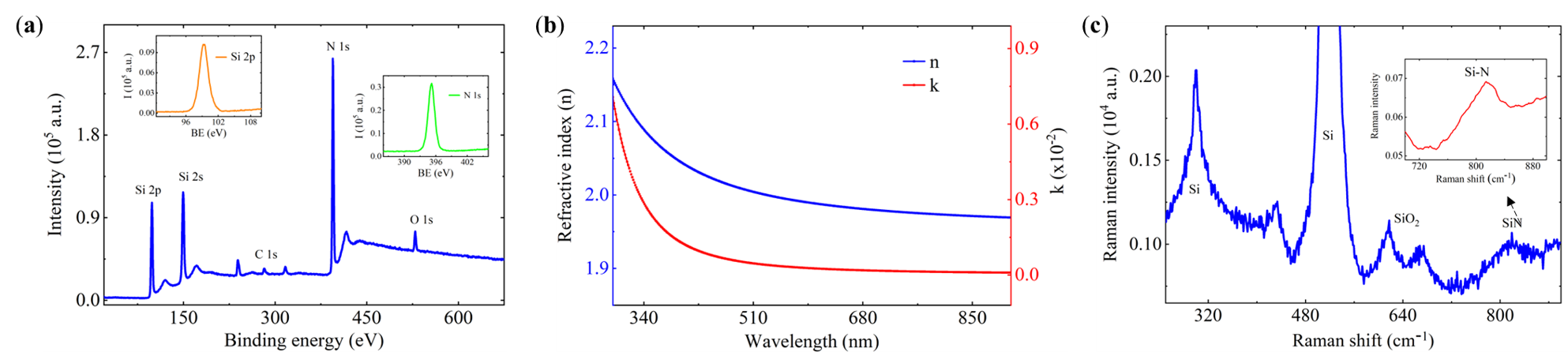}
    {\caption{\textbf{Optical characteristic of as-grown 300-nm-SiN thin layer material via PECVD system.} (a) XPS surface analysis to examine material composition. The inset shows the XPS spectra of elements (Si and N) in their binding energy span. (b) The dispersion (n, k) plot through spectroscopic ellipsometry by employing the Tauc-Lorentz fitting model. (c) Raman spectra of the SiN material vibration mode, peak value centered at 790 cm$^{-1}$ as shown in insets of scan in the vicinity of Raman mode of SiN.}}
\end{figure*}
 To create a notch in the outer edge of 30-$\mu$m-dia SiN microring of 600-nm-width size, we used the subtractive lithography technique in a high-kV FIB/SEM system (Helios 5 UC, Thermoscientific). A subwavelength size notch in a half-disk shape of radius $\lambda$/3 was considered in our work. In a dual-beam FIB/SEM system, initially, we marked the desired half-disk soft mask by e-beam exposure, and subsequently Ga$^{3+}$-ions focused beam was bombarded for drilling with controllable etch rate until reaching the interfacial bottom SiO$_{2}$ layer. In this process, a 0.23 pA collimated beam current (beam size $\sim$ 12.3 nm) was exposed to a selected area on the sample at 30 kV accelerating voltage with a pixel resolution of 6144~x~4096. Further, to eliminate a small fraction of Ga$^{3+}$-ions intercalated in the vicinity of the notch section, the device sample was thermally annealed at 300 $^{\circ}$C for 3 hrs \cite{sahoo2023polaritons}. Fig. 1(d) displays the notch-integrated microring cavity and the insets show magnified view of the same. \\
In the end, the as-fabricated photonic chip was again cleaned with DI water, acetone, IPA, and DI water then dried with nitrogen gun and heated at 200 $^{\circ}$C. Finally, oxygen plasma treatment at 75 W in RIE was carried out to remove residue or dirt particles which may have adhered to the microcavity sidewall surface and at the notch location, before the optical measurements. The atomic force microscopy (AFM) topography is shown in Fig. 1(e).\\

\section{RESULTS AND DISCUSSION}
\subsection{SiN thin film characterization}
\textbf{Chemical composition in XPS analysis}. The presence of elements, composition, and bond state of the as-deposited SiN thin film was validated by X-ray photoelectron spectroscopy (XPS, AXIS Supra from Kratos Analytical, UK) with the irradiation of monochromatic Al K$_{\alpha}$ (of energy $\sim$1486.6 eV) X-ray photons. The surface survey scan (of depth $\sim$10 nm) in XPS from the range of 25-700 eV displays the core-level spectra of constituent elements of the film. The XPS spectrum confirmed the dominance of both the major elements: silicon (Si) and nitrogen (N) in the film with highly intense peaks, as shown in Fig. 2(a). Additionally, a few minor peaks were observed which referred to the existence of carbon (C), oxygen (O), and argon (Ar) elements. The presence of Ar in the film could be due to the use of Ar-plasma in the PECVD chamber. Due to adsorbed ambient gas molecules and also the high reactivity of Si with O, we observe the occurrence of O in the film surface layer. The C element was mainly sample surface contamination in the environment before the XPS measurement. For the identification of the bonding state of Si-N, we investigated the XPS core-level spectra of both elements Si$_{2p}$ and N$_{1s}$, as plotted in insets of Fig. 2(a). We find peaks centered at the binding energy $\sim$100 eV (attributed to Si-N network) and $\sim$ 396 eV (attributed to N) respectively\cite{yao2006fabrication}. The estimated atomic concentration ratio (Si$_{2p}$/N$_{1s}$) is $\sim$ 0.72, which signifies the film composition to be close to the stoichiometric configuration\cite{lin2013low}.\\\\ 
 \textbf{Spectroscopic ellipsometry}. We have performed the spectroscopic ellipsometry (J. A. Woollam tool) to extract the optical constants ($n, \kappa$ being the real and imaginary parts of the refractive index) of SiN thin film of the optimized thickness value (estimated by profilometry scan by creating a step). In the measurement setup, we have taken a stacked layer model (Si/SiN/air) and obtained the ellipsometric angles ($\Psi$ and $\Delta$) data at a 45 $^{\circ}$ angle of incident light polarization, and fitted with the Tauc-Lorentz model (supplementary section S1) \cite{borojevic2016advanced}. The optical dispersion dependence with wavelength is plotted in Fig. 2(b) for an optimized film thickness of $\sim$ 300nm. The observed refractive index value at wavelength 600 nm is $n\sim1.94$.\\\\
\textbf{Raman spectra}. The vibrational response of our SiN material was analyzed through Raman spectroscopy (Invia reflex, Renishaw). In this study, the SiN microring cavity surface was excited with an incident-focused laser light of 532 nm wavelength with an excitation power of 1.8 mW. The Raman spectrum was collected in the backscattered mode with a microscope objective (100x, of NA = 0.90), a spectrometer with 2400 grooves/mm grating, and an acquisition time of 10 sec. The wide band spectrum ranging from 745 cm$^{-1}$ to 845 cm$^{-1}$, and peak centered around $\sim$810 cm$^{-1}$ signify the amorphous nature of SiN material. Also, peak at $\sim$810 $^{-1}$, suggests nano-clusters of Si encapsulated in a SiN material matrix and attributed by asymmetric stretching vibration mode of the Si-N bond state in $N_{2}-Si-H_{2}$ group \cite{bandet1999nitrogen}. Additionally, two peaks belong to Si one at 521 cm$^{-1}$ is to the optical phonon scattering within nano-crystalline Si and at around 302 cm$^{-1}$ correspond to the longitudinal acoustic mode in Si \cite{Mankad2012}. The Raman peaks at around 640 cm$^{-1}$ are from D$_2$ band in the SiO$_{2}$ material \cite{Borowicz2012}.

\begin{figure*} 
   \centering
   \includegraphics[width= 0.99\textwidth]{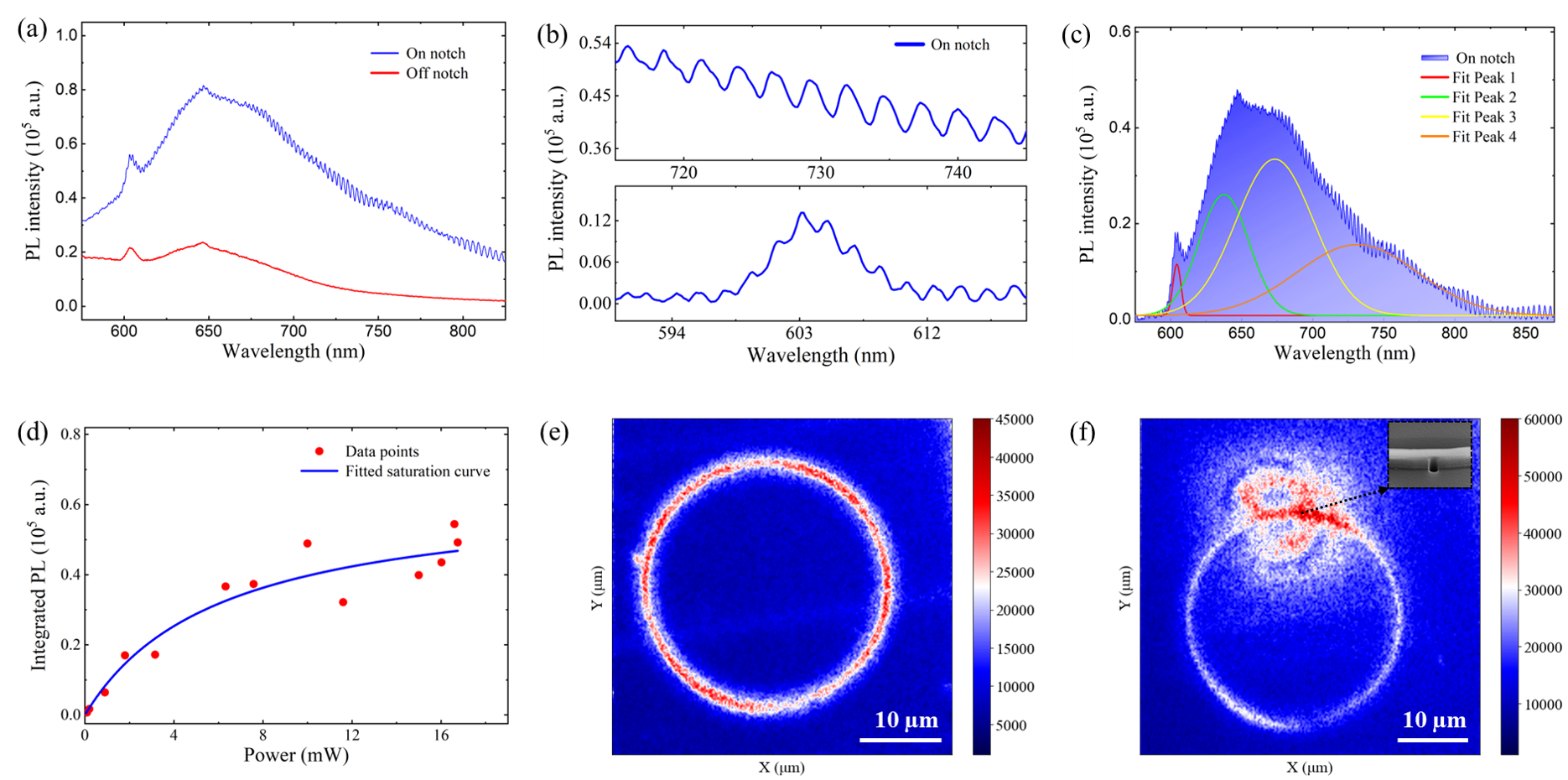}
    {\caption{\textbf {Photophysical properties in a SiN microring cavity platform}. Emission of cavity-coupled PL spectra through excitation of 532 nm laser light at the edge of microring cavity nanophotonic platform in a direct excitation/detection scheme. (a) The PL spectrum was recorded without (red solid line) and with (blue solid line) a notch structure in the microring cavity. (b) Shows the WGM mode signature of the cavity in a narrow wavelength scan at two distinct central peak positions at $\sim$604 nm (bottom panel) and $\sim$730 nm (top panel) respectively. (c) PL spectrum fitted with Gaussian-line shape and deconvoluted into four distinct central peaks, which suggest the origin of distinct types of defect centers in the SiN matrix. (d) Power-dependent PL study and fitted for saturation curve of integrated PL spectra vs input excitation that gives the saturated power at $\sim$6 mW. The conformal PL-intensity mapping of SiN microring resonator without (e) and with (f) notch structure at the rim, inset specify the notch site. Indicates a signature of trapped cavity mode inside the cavity at room temperature.}}
\end{figure*}
\subsection{SiN emitter characterization}
\textbf{Photoluminescence measurement setup: }
After the fabrication process, the photophysical characteristics of inherent emitters inside the SiN microring resonator cavity structure were explored by employing a commericial photoluminescence (PL) setup (Invia reflex, Renishaw). We used a 532~nm continuous wave (CW) solid-state diode laser. The input laser light beam was focused via a microscope objective (Leica, 100x with NA $\sim$0.90), and emitted photons were captured with the same objective from the photonic chip surface. The sample was mounted on a motorized XY-piezo stage to precisely locate specific points on the microring in an excitation under normal incidence. The incoming focused laser beam size was approximately 1~$\mu$m. The spectrometer with a grating of 600 grooves per mm was selected for capturing spatially resolved spectra in the wavelength ranging from 570 to 900 nm. All measurements were carried out in ambient conditions. \\
To conduct confocal photoluminescence (PL) intensity mapping, a customized scanning confocal spectroscopy system was employed on a free-space optical setup. The optical setup included a 40-$\mu$m pinhole and an air microscope objective (Olympus, 100x magnification with NA of 0.90). Optical excitation of emitters was achieved using a continuous-wave 532-nm laser from Thorlabs. The 550-nm long-pass dichroic mirror (DMLP550L, from Thorlabs) was employed to separate the excitation light from the PL signal. A 550-nm wavelength long-pass filter (FEL0550, from Thorlabs) was placed in front of the detectors to filter out the pump beam.\\
\textbf{Emitters spectral characteristics: }
The intrinsic emitters inside silicon nitride material were excited using a $\mu$-PL setup with a 532-nm laser pump beam at 1.8~mW excitation power. As shown in Fig. 3(a), the cavity-coupled PL spectra were captured for a set acquisition time of 30 sec from the SiN microring architecture platform designed with a notch (blue solid line) and without a notch (red solid line) at the edge of microring as depicted in Fig. 1(c-d) insets. This corresponds to a site-specific excitation and detection configuration for free space collection in visible-to-near infrared wavelength regions. The collected PL intensity from a notch region in the cavity was found to have notably higher in intensity compared to those without a notch cavity under identical excitation conditions. This enhancement can be explained on the basis of improved coupling efficiency for both excitation and collection via a notch, as shown in the field plot in Fig. 3(f). A notch acts as an antenna for coupling the excitation beam into the cavity as well as strengthens the signal outcoupling via increased scattering. An unpatterned SiN film on SiO$_{2}$ layer was also tested before defining the cavity structure in fabrication (supporting material section S2) and the recorded PL-intensity in this case was much lower in comparison to the cavity-enhanced PL.\\
Remarkably, the notch-associated SiN-microring cavity nanophotonic configuration displays the resonating modes (whispering-gallery-mode) distributed in the envelope of the cavity-coupled PL spectrum, Fig. 3(b) illustrates the magnified spectrum around two specific wavelengths centered at 600 nm and 725 nm (analyzed in detail in the next section). Additionally, the cavity-PL spectrum exhibits multiple resonances in the deconvoluted emission spectrum and can be well-fitted using four Gaussian-line shapes. As shown in Fig. 3(c), the spectral positions of these peaks are around 600~nm, 650~nm, 680~nm, and 750~nm. This fitting was carried out after appropriate baseline correction of the spectra. Here, the most prominent peaks are identified at 650~nm and 680~nm, whereas the remaining two peaks lie at the shoulder. We have observed similar signatures of all four peaks in multiple samples fabricated in the same as well as different batches using the same recipe. This multipeak feature suggests several kinds of embedded emitters or defect families in the SiN matrix\cite{senichev2021room}, which requires a more in-depth exploration to understand their origin and is beyond the scope of the current work. However, for the cavity without the notch, only two peaks were observed instead of four. The exact reason for this is unclear however it might be possible that the excitation of the other two modes might be suppressed due to weak coupling (supplementary material S3 for spectra). \\
The excitation power-dependent investigation was conducted to examine the behavior of the L-L curve (incident laser pump power as a function of the integrated output power of emission peak, representing light-light plot) of the notch integrated cavity-coupled PL spectrum of a selected peak at 600~nm wavelength. Fig. 3(e) portrays the estimated integrated power (of a de-convoluted peak at ~600 nm) as a function of experimental input laser power (in linear scale). The data points were fitted by an equation of the form:
\begin{equation}
I(P)=I_{\infty} \times P /\left(P+P_{\text {sat }}\right)
\end{equation} 
where $I_{\infty}$ and $P_{\text {sat }}$ are fitting parameters denoting maximum intensity and saturated power components respectively. The fitted PL saturated curve yielding parametric values are  $I_{\infty}$ $\sim$6x10$^{4}$ cps and  $P_{\text {sat }}$ $\sim$6~mW.\\
For the demonstration of field confinement inside the microring cavity, the far-field confocal PL intensity map was obtained by using a charged coupled device (CCD) camera in both the 30-$\mu$m-dia SiN microring cavity configurations, that is, without (Fig. 3(d)) and with (Fig. 3(f)) the notch. In the microring with no notch, the spatially resolved cavity PL map indicates modes that are tightly confined and re-circulated via ``reflection" at the interface of the resonator sidewalls. Whereas in the microring with a notch, the cavity-PL from the emitters becomes dominant at the notch site, in addition to showing the inevitable scattering losses through the waveguide sidewall. The notch therefore leaks out a substantial portion of the cavity mode with significantly higher counts compared to the case without the notch. This occurs due to the geometry of the intentional defect and is well justified via a simulated field profile, as shown in Fig. 4(d) and supplementary section S2. \\

\subsection{Theory: WGM cavity modes in microring}

The proposed nanophotonic chip shows the cavity-enhanced WGM resonating modes in $\mu$-PL spectrum through localized emission at the notch site. The spatial field distribution ($\mathbf{\psi}=\{\mathbf{E}, \mathbf{H}\}$) with distinct mode numbers in the microcavity are the solutions of classical Helmholtz wave equation: $ \left(\nabla^2+k^2\right) \psi=0 $, where the wavevector ($k = k_{n,m}$) is $n_{\mathrm{eff}}^2 \omega^2 / c^2$, $\omega=2 \pi c / \lambda_{n,m}$ and $n_{\mathrm{eff}}$ denotes effective index of mode. Under the approximation of a 2D-microring planar system, we consider the solution in polar coordinates (r-$\phi$ plane) with field vector of form $\psi_{n,m} (r, \phi)$ = $\Re(r)~\Phi(\phi)$, where the radial and azimuthal mode numbers are labeled with n, m respectively, which are the characteristics of confined mode. The resonating mode $(n, m)$ field distribution inside the microring ($R-w/2 \leq r \leq R+w/2 $, $w$ is strip waveguide width) can be written as \cite{van2016optical, hansson2014analytical, heebner2008optical};
\begin{figure}
   \centering
   \includegraphics[width= 0.45\textwidth]{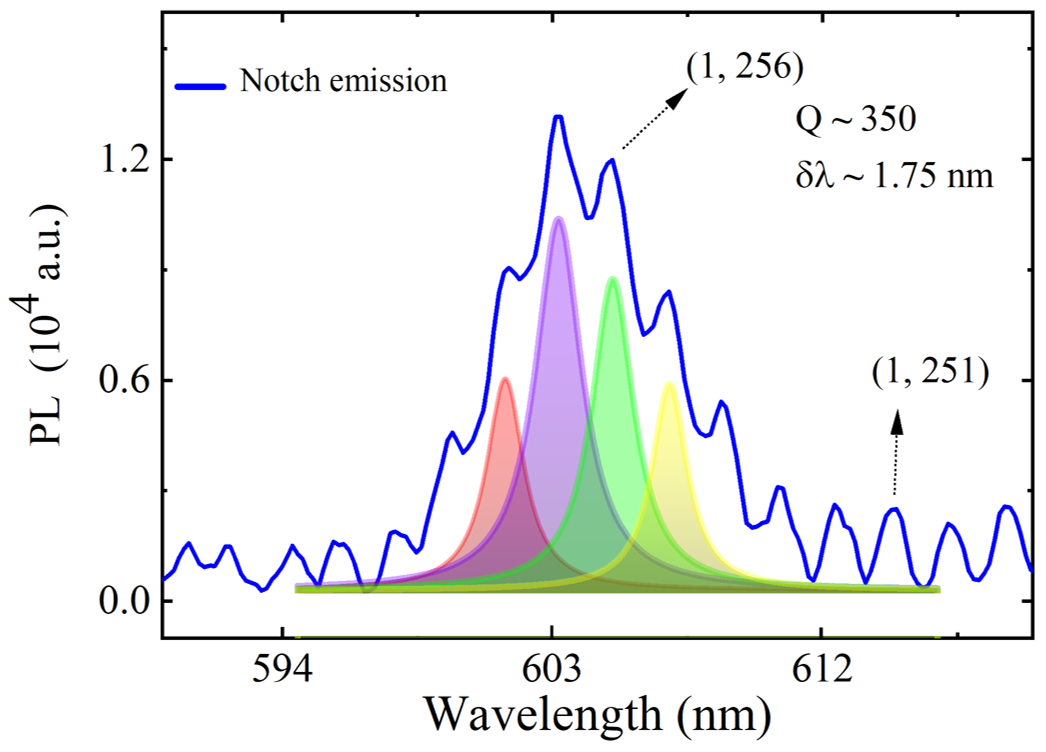}
    {\caption{Modal analysis of the spectral distribution of WGM cavity mode exhibited by SiN microring resonator at a localized notch site. The shaded area represents the deconvoluted peaks of the fitted Lorentzian line shape.}}
\end{figure}
\begin{equation}
\psi_{n,m} (r, \phi) \approx [\alpha~J_n\left(k r\right)+ \beta~Y_n\left( k r\right)] ~e^{ \pm i m \phi}
\end{equation}
where $J_m\left(k r\right)$ is the Bessel function, $Y_m\left( k r\right)$ denotes the Neumann function and $\alpha$, $\beta$ are the constant coefficients. In the case of microdisk, the WGM modes are confined strongly between the inner caustic radius (strong optical inertia blocks penetration of radial field) and outer radius interface \cite{heebner2008optical}. 
\begin{figure*}
   \centering
   \includegraphics[width= 0.9 \textwidth]{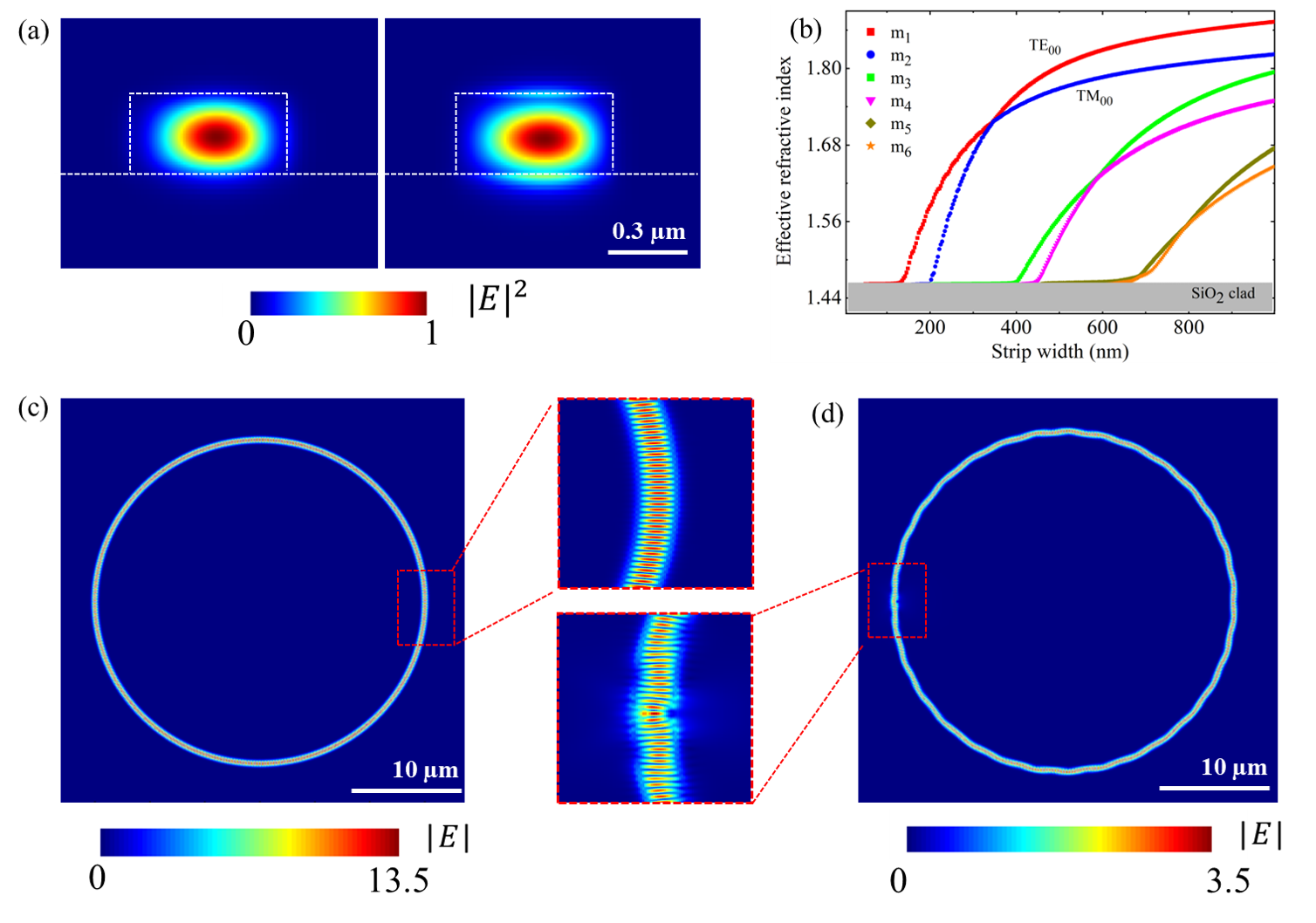}
    {\caption{\textbf{Engineering the resonating mode by simulating microring architecture using Ansys/Lumerical software.} (a-b) Modal analysis of bent waveguide of strip type cross-section to satisfy the single mode condition in microring resonator via FDE solver. (b) The effective refractive index variation with strip width (w) sweeps at the central radius by keeping height fixed at $h\sim$300 nm. The contour plot of the electric field profile of fundamental TE-like mode (left panel) and TM-like mode (right panel) in (a) supported in the microring cross-section. The FDTD-simulated electric field distribution inside the microring cavity without (c) and with (d) a sub-wavelength side notch scatter at the rim. The insets show the zoomed-in view of confined WGM cavity mode.}}
    \label{fig:figure6}
\end{figure*}

We investigate the spectral characteristics of the WGM mode in our microring cavity around the emission wavelength of 600 nm. In the WGM spectrum, the resonating modes belonging to a particular mode family are equally spaced and this spacing is known as the free-spectral range (FSR). The FSR is estimated by measuring the peak separation ($\Delta \lambda \approx \lambda_{m+1} - \lambda_{m}$) and defined by the equation \cite{heebner2008optical}; 
$$
\mathrm{FSR}= \Delta \lambda = \frac{\lambda^2}{2 \pi R~n_{\mathrm{eff}}}
$$
The experimentally measured average FSR value is close to $\Delta \lambda_{PL}\sim 2~nm$, which is well matched with the theoretically calculated value $\Delta \lambda_{sim} \sim 2.08 ~nm$. Another important key parameter of WGM types cavity is the quality factor $(Q)$ which is a measure of the ability of the resonator to trap photons. A high $Q$ means that the cavity mode is long-lived and is directly related to the sharpness of the resonant cavity mode:
$$
Q= \frac{\lambda}{\delta \lambda}
$$
where $\delta \lambda$ is the linewidth of the resonant peak and is calculated by full-width at half maxima (FWHM). We have fitted the PL spectra with Lorentzian lineshapes as shown in Fig. 4 and extracted an average linewidth of $\delta \lambda \sim$1.75 nm, which gives an average value of the loaded Q-factor of about $\sim$350. The finesse ($\mathcal{F}$) measures the microcavity lifetime in the context of the round trip time of resonating mode and is calculated as $\mathcal{F} \sim  \frac{\Delta \lambda}{\delta \lambda} \approx Q/m$ and the estimated value is about 1. \\
In general, the Q-factor is computed by summation of all types of loss factors associated with a microcavity \cite{yang2015advances, borselli2005beyond, kekatpure2008fundamental}, $$Q^{-1}=\sum_j Q_j^{-1}$$ $Q_j$ includes scattering loss channels as well as material absorption channels, emanating from the shape and size of the microcavity, material deposition conditions,  and fabrication imperfections. These losses overall affect the device's performance and reduce the Q-value. \\

The material loss term depends on the optical absorption of the SiN material. SiN typically has a very small imaginary part of the refractive index (refer to Fig. 2(b)), which results in the absorption coefficient value of $\alpha_{abs}~(\approx 4\pi \kappa/\lambda) \sim\text{40~cm}^{-1}$ at a wavelength of 600 nm. The ultimate absorption limited Q-factor is approximated with \cite{xu2009influence},
$$
Q_{abs}= \frac{2\pi ~n_{eff}}{\lambda ~\alpha_{abs}} \ge 10^{3}
$$
Material absorption losses can be reduced by minimizing the $k$ value in the growth process.\\ 
We note that scattering loss limited Q-factor mainly originates due to film surface non-uniformity, side-wall edge roughness during the lithography and etching process, oxidation of the top surface and adhesion of contamination particles in open air ambient during measurement as can be seen in Fig. 1(c-e) and Fig. 2(a). While fabrication imperfections are inevitable, however, this can be minimized by employing post-processing approaches on device samples such as chemical polishing for smoothing surfaces, annealing, and testing under a clean-room environment \cite{brenner20173d}. The patterned and un-patterned SiN surface roughness was measured with AFM and has an average value of $\sim$6.25 nm (measured after optical measurement) and $\sim$1.53 nm respectively, as shown in Fig. 1(e) and supplementary section S3.\\
Importantly, the radiative losses of resonating mode in WGM microcavities are specifically morphology-dependent; where the index contrast between core and cladding material and shape or size of the radius of curvature determined the minimal bending loss of re-circulation of the cavity mode in small mode volume. As these parameters decrease, they contribute to a significant increase in radiation loss and decrease in the $Q_{rad}$, as the coupled photon is weakly confined inside the microcavity. $Q_{rad}$ is a function of mode numbers $(n,~m)$, which signifies a low $Q_{rad}$ factor for higher order modes in comparison to the fundamental mode.

\subsection{Simulation: photonic chip design}
\textbf{Modal analysis of SiN microring in FDE: } 
Single-mode waveguide condition is typically optimal for use in integrated photonics. Strip-type waveguide is taken into consideration for making a bent waveguide structure, offering flexibility with substantial mode confinement. We have performed FDE simulations to optimize the bent waveguide cross-section dimension (height and width, $h \times w$) by plotting the effective index versus dimensions $(h, w)$ \cite{kumar2023photonic}. In the simulation model, we have specified material parameters (2-$\mu$m-thick SiO$_{2}$ as a clad layer of $n\sim$1.45 and SiN quantum emitter core layer of $n\sim$1.94). Firstly, the SiN layer thickness was varied in the 1D-FDE setup and we obtained a $\sim$ 300~nm cut-off value for single-mode operation in planar film, (see supplementary section S5).\\
The bending radius $R$ also plays a pivotal role in cavity mode propagation losses. A 30-$\mu$m-diameter SiN-microring resonator taken into account with an optimal bending loss of propagating cavity-coupled emitter mode. At a fixed value of bend radius (R$\sim$15$~\mu$m) and thickness (h $\sim$300 nm), the microring width was swept around the center of the radius of curvature with appropriate bend orientation. The FDE simulation region extended up to nearly $\sim$4$\lambda$ from the strip width center. The computed cut-off value of width was observed around $\sim$410 nm and is marked as a transition boundary line between fundamental modes and higher order mode family in the waveguide cross-section dimension in Fig. 5(b). The achieved single-mode operational waveguide is found to have maximum dimensions of $300 \times 410$ nm$^{2} (h x w)$. All the optimization processes were carried out for the quantum emitter emission wavelength peak at 600 nm, as discussed in the previous section. \\
In the experimental setup of $\mu$-PL, the minimum spot size that can be achieved with the focused laser beam of approximately 1~$\mu$m. We have therefore chosen a slightly increased value of waveguide strip width to 600~nm (average value of as-fabricated sample) for efficient coupling-in/out of light signal in an out-of-plane coupling scheme on integrated photonic chip platform, due to limitations of minimum spot size in the present setup. A 600-nm waveguide width (at 300-nm height) supports up to two mode orders (0$^{th}$ and 1$^{st}$) and works as a multimode waveguide at $\sim$600 nm (see supplementary section 6). \\
Thus in our simulation, we mimic the experimental configuration and consider a SiN-microring strip cross-section dimensions as $300 \times 600$ nm$^{2}$ (h x w). The supported fundamental mode field profile of TE (left side) and TM-like polarised mode (right side) are shown in Fig. 5(a) and the 1$^{st}$ order TE and TM-type mode field map are depicted in supplementary section S4. \\\\
\textbf{FDTD simulation of WGM cavity resonance in microring:} Experimentally, we have demonstrated the coupling dynamics of intrinsic emitters into the SiN microring resonator platform and out-coupled these emitter cavity modes via notch engineering within the same SiN material system. To visualize the steady-state resonating modes in our proposed nanophotonic platform, we have simulated the SiN-microring photonic chip using an FDTD solver. In the FDTD simulation setup, we have taken the same device parameters as discussed in the previous section and used a perfectly matched layer (PML) boundary condition at $\sim$4$\lambda$ distance from microring. An emitter in the SiN-microring material matrix was modeled by placing an electric dipole source inside the microring (positioned at the center of the strip waveguide cross-section) to excite the cavity mode. This electric dipole dominantly excites the transverse electric (TE)-like cavity modes.  In a high-Q cavity, such as a SiN microring, the electric field persists due to a long-lived resonating mode, resulting in slow decay of the time domain signal and not fully decaying to zero even with increased simulation time because of source excitation. To capture the steady-state response of the microcavity resonating mode spectrum, we used the apodization function in the frequency domain monitor to extract the cavity response spectrum in a particular simulation time window where the signal was forcefully terminated to zero by the end of the simulation \cite{wong2021epitaxially}.\\
We have simulated the microring structure by including and excluding a notch design with a dipole source around the central wavelength peak at 600 nm. 
In a microring without notch, the spatially separated WGM cavity mode depicted in the simulated electric field ($|E|$) distribution profile in Fig. 5(c), inset shows the enlarged view of the coupled resonating mode at wavelength $\sim$604.5~nm. Similarly, by introducing a notch of size ~$\lambda/3$ (shown in the index monitor map, supplementary section 7), a WGM cavity mode response was obtained, as illustrated in Fig. 5(d). A notch distorted the pattern of standing-wave cavity mode distribution as observed in the absence of a notch, as portrayed in the inset, which may be caused by a sudden interruption at the notch section. At a notch site, trapped WGM resonating modes are enforced to leak out via scattering loss and make them easily accessible for collection vertically (shown in a zoomed-in view of the mode profile in linear Fig. 5(d) inset and logarithmic scale, in the supplementary section 8).  \\  
\section*{Conclusion}
The exceptional optoelectronic properties of wide band-gap silicon nitride (SiN) are intriguing as a photonic technology that fosters a large-scale on-chip integration with promising device flexibility, and functionality in the visible to infrared wavelength regime. However, the hybrid integrated photonic chip system increases fabrication complexity and leads to degradation of optical coupling and losses. Seeking a CMOS-compatible optical material with dual functionality, serving as an intrinsic light source as well as a host platform for the photonic element will mitigate these issues. In this work, we have demonstrated such a platform by growing the SiN material in a PECVD system which manifests an intrinsic emitter and subsequently coupling these emitters into a microring-resonator nanophotonic structure made of the same SiN material. We demonstrated WGMs through a cavity-enhanced PL spectrum by making them accessible via the engineering of a sub-wavelength notch ($\sim \lambda/3$) at the rim of the microring that enables efficient coupling-in/out of light. Our work indicates that silicon nitride (SiN) has the potential to serve as a monolithic integrated photonic platform for nanoscale optical sources. This makes it a potential candidate for on-chip lasing which can complement existing gain media such as group  III-V compounds and heterogeneously integrated gain systems such as two-dimensional atomically thin Van der Waal semiconductors, to name a few -- which are again limited by either CMOS non-compatibility, degradation with hybrid integration and fabrication challenges. Our emission performance was limited by fabrication imperfections leading to modest $Q$ values. Engineering the material growth, cavity design, and reduction in associated fabrication imperfections along with sample post-processing could lead to further improvement in the performance. This discovery extends the scope for the realization of whispering gallery mode based device applications and paves the way for next-generation monolithic nano/microlaser, and quantum photonics technology development on the SiN platform.

\textbf{\emph{Acknowledgment--}}
K.K.M. acknowledges the GATE fellowship No.~-- RSPHD0006 from IIT Bombay, India. A.K.S. acknowledges the IRCC IIT Bombay for funding support. A.K. acknowledges funding support from the Department of Science and Technology via the grants: SB/S2/RJN-110/2017, ECR/2018/001485, and DST/NM/NS-2018/49.
We also acknowledge the Industrial Research and Consultancy Center (IRCC); the Centre of Excellence in Nanoelectronics (CEN), IIT Bombay; Fundamental Optics, THz and Optical Nanostructures (FOTON) laboratory at TIFR-Colaba, Mumbai; and Sophisticated Analytical Instrument Facility (SAIF); Centre for Research in Nanotechnology and Science (CRNTS)  IIT Bombay for providing access in fabricating photonic chip, sample characterization. 
\bibliography{main-ref}
\end{document}